\documentclass[journal]{IEEEtran}
\usepackage{amsfonts}
\usepackage{amsmath}
\usepackage{amsthm}
\usepackage{amssymb}
\usepackage{mathtools}
\usepackage{dsfont}
\usepackage{graphicx}
\usepackage{epstopdf}
\usepackage{cite}
\usepackage{bm}
\usepackage{tikz}
\usepackage{hhline}
\usepackage{multirow}
\usepackage{pgfplots}
\usepackage{enumerate}

\usepackage{epstopdf}
\usepackage{color}

\usepackage{hyperref}
\usepackage{balance}

\theoremstyle{remark} \newtheorem{lemma}{Lemma}
\theoremstyle{remark} 
\theoremstyle{remark} 
\theoremstyle{remark} 
\theoremstyle{remark} 
\theoremstyle{remark}

\ifCLASSINFOpdf
\else
\fi
\begin{document}
%
\title{Fault Diagnosis and Bad Data Detection of Power Transmission Network---A Time Domain Approach}
%
%
%

	\author{Zhenyu~Tan,~\IEEEmembership{Student~Member,~IEEE,} 
		\thanks{The first and the fourth author is with the School of Electrical and Computer Engineering, Georgia Institute of Technology, Atlanta, GA, 30332, USA (e-mail: ztan@gatech.edu).}
		Yu~Liu,~\IEEEmembership{Member,~IEEE,}
		\thanks{The second author is with the School of Information Science and Technology, ShanghaiTech University, Shanghai, 201210, China (corresponding author, e-mail: liuyu@shanghaitech.edu.cn).}
		Hongbo~Sun,~\IEEEmembership{Senior~Member,~IEEE,}
		\thanks{The third author is with Mitsubishi Electric Research Laboratory, Cambridge, MA}
		Bai~Cui,~\IEEEmembership{Student~Member,~IEEE}
}

\markboth{}%
{Shell \MakeLowercase{\textit{et al.}}: Bare Demo of IEEEtran.cls for Journals}

\maketitle

\begin{abstract}
Fault analysis and bad data are often processed in separate manners. In this paper it is proved that fault as well as bad current measurement data can be modeled as control failure for the power transmission network and any fault on the transmission line can be treated as multiple bad data. Subsequently a linear observer theory is designed in order to identify the fault type and bad data simultaneously. The state space model based observer theory allows a particular failure mode manifest itself as residual which remains in a fixed direction. Moreover coordinate transformation is performed to allow the residual for each failure mode to generate specific geometry characteristic in separate output dimensions. The design approach based on the observer theory is presented in this paper. The design allows 1) bad data detection for current measurement, and 2) fault location, and fault resistance estimation (as a byproduct) where the fault location accuracy is not affected by fault resistance.  However it loses freedom in designing the eigenvalues in the excessive subspace. While the theoretical framework is general, the analysis and design are dedicated to transmission lines.
\end{abstract}

\begin{IEEEkeywords}
Fault diagnosis, Luenberger observer, detection subspace, excessive subspace.
\end{IEEEkeywords}

%
\IEEEpeerreviewmaketitle

\section{Introduction}
%
%
%
%
\IEEEPARstart{M}{odern} digital transmission line protective relays are often equipped with individual functionalities including 1) fault detection, 2) single pole tripping, and 3) fault location functions. The single pole tripping are based on use of the phase selector to identify type of the fault, to eliminate incorrect and insensitive fault identification that can be made by distance protection element, and provide trip initiation from elements that are not capable of any fault type identification, such as high-set negative-sequence directional overcurrent element.

The \emph{conventional relay fault identification method} often uses the variations in system voltages, currents, power and frequency to detect fault occurrence, and uses phase relations between sequence current components for fault identification \cite{1,2,3,4,5}. However the settings rely heavily on system level fault study and need to change adaptively with various system conditions. The \emph{conventional fault location methods} with measurements from both terminals of the line are mainly categorized into phasor based method \cite{6}. However the accuracy is often limited with high impedance fault and location of the fault. 

Other theoretical approaches exist as well. The \emph{wavelet transformation method} is a time domain approach that detects high frequency components contained in a fault signal spectrum \cite{7}-\cite{8}. The relays at both terminals of the line computes the time difference between receipt of the high frequency signals. However this method is limited by the data sampling rate during transients at fault inception, which is often in order of MHZ for an accurate estimation of fault location. The artificial intelligence based method utilizes prior knowledge of system quantities (such as voltage and current) under different fault and operating conditions. This knowledge is used to train a learning system to identify abnormal conditions and classify them. Methods such as neural network \cite{9}, fuzzy theory \cite{10}, decision tree \cite{11}, clustering \cite{12} have been developed. The major limitation is heavy computation time that may cause operational delay and system instability.

The model based methods have also been widely investigated. The approach is based on the generation of residual signals using observer theory that reflects the difference between the actual and estimated values of the output as an indicator of fault occurrence \cite{13,14,15,16,17,18}. However this approach does not allow fault identification and location, the main limitation is that the residual being generated has no mathematical meaning. In order to identify the fault type, a multiple model filtering technique is proposed by \cite{19}, given the input and output of the system, the goal is to determine which model, in a set of predefined models, best approximates the actual system behavior. The observer theory has also been adapted for fault diagnosis in other engineering systems where the faults are modeled as control failure \cite{20}, \cite{21}, the feedback in the observer is designed specifically to allow a number of control failures to generate uni-directional residuals. Inspired from \cite{20}, this paper provides a comprehensive approach for the fault diagnosis for transmission lines including fault detection, identification and location. It first models the transmission line under fault condition as control failure, then it provides two different approaches for the fault diagnosis. While the same theory has been applied similarly in converter systems \cite{21}, such focus was on simple degradation in individual components. This paper is dedicated to the transmission network where the faults are much more complicated since it can involve the ground and multiple phases. The originality of this paper include: 1) it proves that the line faults can be treated as control failure through mathematical arrangement, 2) it proves that through specific filter design, the generated residual can have mathematical meaning and the fault location can be derived from the residuals, along with the fault resistance, compared to traditional methods where the fault location is insensitive with high impedance fault.

This paper is organized as follow. In Section \ref{line_model} the transmission line models are described by state space model where phase and ground faults are modeled as control failures. The failure detection/observer theory is included in Section \ref{fail_detec}. The design based on observer theory and \cite{20} is presented in Section \ref{filter}. The proposed research work is studied and compared with state-of-art double terminal fault detection method in \ref{simulation} 

This time domain approach is suited for protective relaying applications where data samples are transmitted among the relays at both terminals of the transmission line.

\section{Transmission Line Models}
\label{line_model}
\subsection{short transmission line}
The transmission line $\pi$-model was specifically developed for fundamental frequency analysis. In this paper the same transmission line model (single-section) is also used for differential-algebraic equations (DAE) modeling as shown in Fig. \ref{fig1} and mathematically formulated below:
\begin{subequations}
	\begin{align}
		i_1(t) &= C_{ap} \frac{\mathrm{d} v_1(t)}{\mathrm{d}t} + i_L(t), \\
		i_2(t) &= C_{ap} \frac{\mathrm{d} v_2(t)}{\mathrm{d}t} - i_L(t), \\
		0 &= -v_1(t) + v_2(t) + Ri_L(t) + L\frac{\mathrm{d}i_L(t)}{\mathrm{d}t},
	\end{align}
	\label{dae_orig}
\end{subequations}
where $C_{ap}, R, L \in \mathbb{R}^{4 \times 4}$ are capacitance, resistance and inductance matrices defined as $Cap = [C_{ij}], R = [R_{ij}], L = [L_{ij}]$. 

$i_1(t), i_2(t), v_1(t), v_2(t), i_L(t) \in \mathbb{R}^4$ are state vectors, for example, 
$
	i_1(t) = \begin{bmatrix}
			i_{a1}(t) & i_{b1}(t) & i_{c1}(t) & i_{n1}(t)
		\end{bmatrix}^T.
$ 


\begin{figure}[!t]
	\centering
	\includegraphics[width=3.5in]{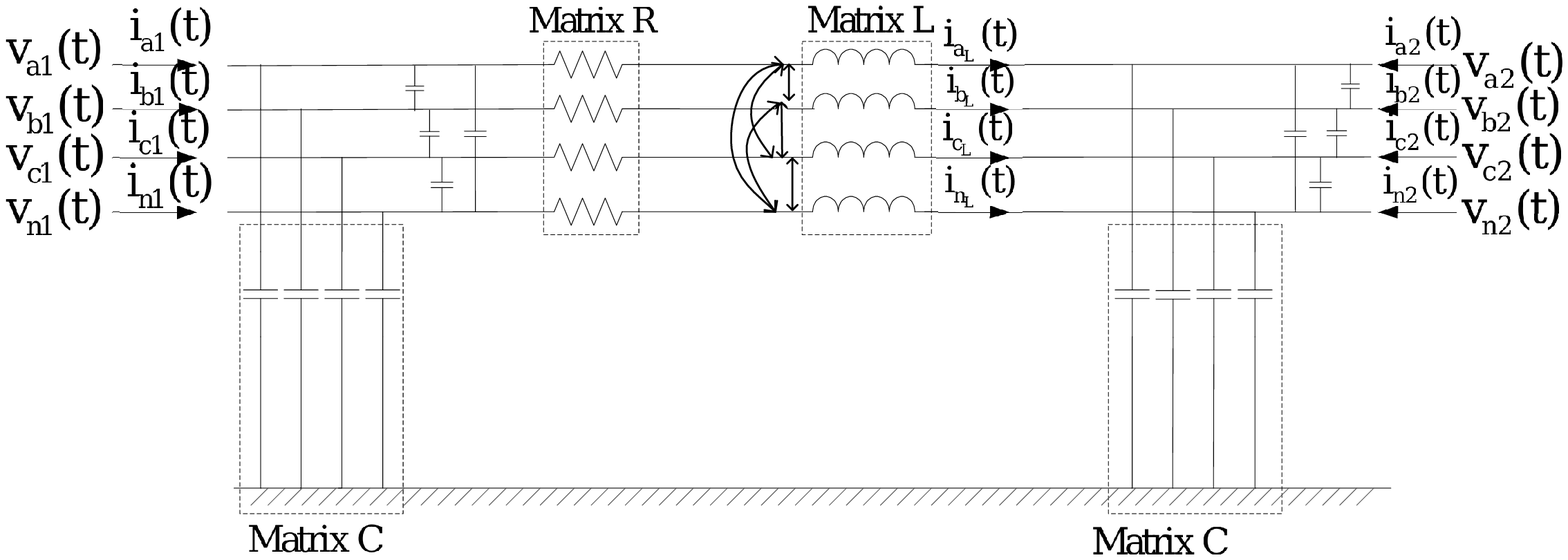}
	\caption{transmission line model -- normal operation}
	\label{fig1}
\end{figure}

\begin{figure}[!t]
	\centering
	\includegraphics[width=3.5in]{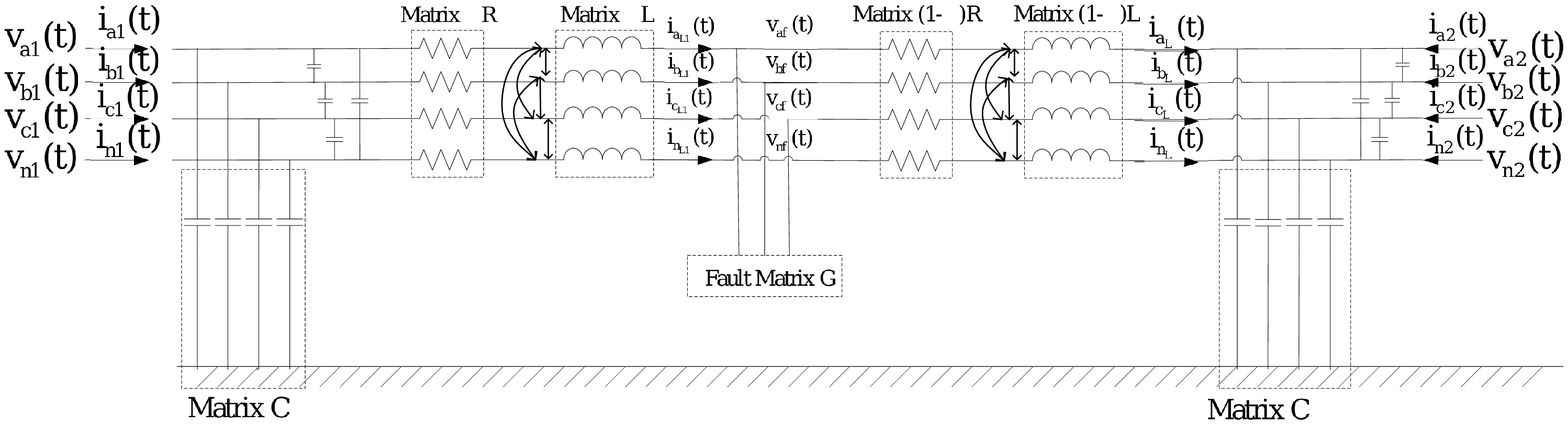}
	\caption{transmission line mode -- fault condition.}
	\label{fig2}
\end{figure}

When a fault occurs, the transmission line model (\ref{dae_orig}) changes to
\begin{subequations}
	\begin{align}
		i_1(t) &= C_{ap} \frac{\mathrm{d} v_1(t)}{\mathrm{d}t} + i_L(t), \\
		i_2(t) &= C_{ap} \frac{\mathrm{d} v_2(t)}{\mathrm{d}t} - i_L(t), \\
		0 &= -v_1(t) + v_f(t) + \alpha Ri_{L1}(t) + \alpha L\frac{\mathrm{d}i_{L1}(t)}{\mathrm{d}t}, \label{dae_fault_3}\\
		0 &= -v_f(t) + v_2(t) + (1-\alpha) Ri_{L2}(t) + (1-\alpha) L\frac{\mathrm{d}i_{L2}(t)}{\mathrm{d}t},  \label{dae_fault_4}\\
		0 &= -i_{L1}(t) + i_{L2}(t) + Gv_f(t).
	\end{align}
	\label{dae_fault}
\end{subequations}
where $\alpha$ is the fault location with respect to the left terminal $0<\alpha<1$, $v_f(t)$ is the voltage vector at fault location, $i_{L1}(t)$ is the inductor current vector to the left of the fault, and $i_{L2}(t)$ is the inductor current vector to the right of the fault, as shown in Fig. \ref{fig2}.

The fault matrix $G$ is dependent on the specific fault type, for example a phase A to ground fault derives the following fault matrix:
\begin{equation}
G = \frac{1}{R_f}
\begin{bmatrix}
1 & 0 & 0 & 0 \\
0 & 0 & 0 & 0 \\
0 & 0 & 0 & 0 \\
0 & 0 & 0 & 0 \\
\end{bmatrix}.
\end{equation}

%

The equations in (\ref{dae_fault}) show that the transmission line fault has to be modeled by augmenting the state space which includes the inductor current on both sides of the fault, and the voltage at the fault location. The fault model in (\ref{dae_fault}) is loosely coupled from the model in (\ref{dae_orig}). In order to strongly couple the two models, the following mathematical treatment is needed.

By summing (\ref{dae_fault_3}) and (\ref{dae_fault_4}) and defining a new state as
\begin{equation}
i_L'(t) = \alpha i_{L1}(t) + (1-\alpha) i_{L2}(t),
\end{equation}
the KVL in fault condition is described by the following equation:
\begin{equation}
0 = -v_1(t) + v_2(t) + Ri_L'(t) + L\frac{\mathrm{d}i_L'(t)}{\mathrm{d}t}
\end{equation}

Note that during normal condition $i_L'(t)$ is the same as $i_L(t)$. Thus (\ref{dae_fault}) can be written in the same form as (\ref{dae_orig}) after mathematical rearrangement:
\begin{subequations}
	\begin{align}
		i_1(t) &= C_{ap} \frac{\mathrm{d} v_1(t)}{\mathrm{d}t} + i_L'(t) + (1-\alpha)Gv_f(t), \\
		i_2(t) &= C_{ap} \frac{\mathrm{d} v_2(t)}{\mathrm{d}t} - i_L'(t) + \alpha Gv_f(t), \\
		0 &= -v_1(t) + v_2(t) + Ri_L'(t) + L\frac{\mathrm{d}i_L'(t)}{\mathrm{d}t}.
	\end{align}
\end{subequations}

The above model can be further translated into the state space model:
\begin{equation}
\frac{\mathrm{d}B_1x(t)}{\mathrm{d}t} = -\left( A_1B_1^{-1} \right)(B_1x(t)) + 
\begin{bmatrix}
i_1(t) \\
i_2(t) \\
0
\end{bmatrix} - f(t),
\label{int_fault_ssm}
\end{equation} 
where
$
f(t) = \begin{bmatrix}
(1-\alpha)Gv_f(t) & \alpha Gv_f(t) & 0
\end{bmatrix}^T.
$ 
And the measurement vector $y$ include three phase to neutral voltage measurements on both terminals, and neutral to ground voltage measurements on both terminals (pseudo measurements).
\begin{equation}
y(t) = \left(C B_1^{-1}\right) (B_1x(t)),
\end{equation}
\begin{equation}
C = 
\begin{bmatrix}
K & 0_{4\times4} & 0_{4\times4} \\
0_{4\times4} & K & 0_{4\times4} \\
\end{bmatrix}
\end{equation}
\begin{equation}
K = 
\begin{bmatrix}
1 & 0 & 0 & -1 \\
0 & 1 & 0 & -1 \\
0 & 0 & 1 & -1 \\
0 & 0 & 0 & 1 \\
\end{bmatrix}
\end{equation}

where $x(t) = \left[ v_1^T(t), v_2^T(t), i_L^T(t) \right]^T$. The vector  $u(t) = \left[ i_1^T(t), i_2^T(t), 0^T(t) \right]^T$ acts as the control input for the system. The fault residual $f(t)$ can be derived separately for single phase to ground faults, phase to phase faults, three phase fault, phase current channel bad data (left or right terminal), phase voltage channel bad data (left or right terminal). Examples for single phase A to ground fault and current bad data on left terminal are:
\begin{subequations}
	\begin{align}
	f_{\textnormal{A-G}}(t) &= \begin{bmatrix}
	(1-\alpha) & 0_{1 \times 3} & \alpha & 0_{1 \times 7}
	\end{bmatrix}^T \frac{v_{af}(t)}{R_f}, \label{fault_event_1} \\
	f_{\textnormal{bad,ia1}}(t) &= \begin{bmatrix}
	1 & 0_{1 \times 11}
	\end{bmatrix}^T i_{a1}(t), \label{fault_event_2}
	\end{align}
	\label{fault_event}
\end{subequations}
where $v_{af}(t)$ is the phase A to ground voltages at the fault location, $f$ is named the fault magnitude function. Here the constant vector (as a function of $\alpha$) is named fault event vector. f(t) with subscript is named fault residual. By comparing \ref{fault_event_1} and \ref{fault_event_2}, the fault event vector for single phase A to ground fault is simply a linear combination of the fault event vectors for phase A bad current measurements (left and right terminal).

The states are left multiplied by $B_1$ for simple definition of fault event vectors.

In general, the transmission line model in fault condition can be summarized as an additional term as the model in normal condition:
\begin{subequations}
	\begin{align}
	\frac{\mathrm{d}x(t)}{\mathrm{d}t} &= Ax(t) + Bu(t) - f(t), \\
	y(t) &= Cx(t).
	\end{align}
\end{subequations}

\subsection{long transmission line}
The single-section model is well-suited for short transmission lines, e.g., 20 miles, while long transmission lines can be modeled by concatenating multiple single-section $\pi$-models.

\section{Failure Detection Theory}
\label{fail_detec}

The principal concern of failure detection is to combine the state estimation capability with a failure detection capability.

The Luenberger observer is to estimate the states by the process model and the output feedback:
\begin{equation}
\dot{\hat{x}}(t) = A \hat{x}(t) + Bu(t) + D(y(t) - C\hat{x}(t))
\end{equation}

When the reference model accurately represents the system, the estimation error and output error can be determined in terms of the error equations:
\begin{subequations}
	\begin{align}
	\hat{\varepsilon}(t) &= (A - DC)\varepsilon(t) \\
	\varepsilon'(t) &= C\varepsilon(t)
	\end{align}
\end{subequations}
where $\varepsilon(t) = x(t) - \hat{x(t)}$ is defined as the estimation error and $\varepsilon'(t)$ is defined as the output error. For the reference model to be an asymptotically stable state estimator, it is necessary that $D$ be chosen such that all the eigenvalues of $A-DC$ have negative real parts. 

In the event of a transmission line failure, the actual model is equivalent as a controller failure as proposed in Section \ref{line_model}:
\begin{equation}
\dot{x}(t) = Ax(t) + Bu(t) + f_1n_1(t)
\end{equation}

$f_1$ is the time-invariant fault event vector and $n_1(t)$ is the time-varying fault magnitude function.

In the case of fault condition the estimation error can be determined in terms of another error equation:
\begin{equation}
\dot{\varepsilon}(t) = (A - DC)\varepsilon(t) + f_1n_1(t)
\end{equation}

The time functions $\varepsilon(t)$ and $\varepsilon'(t)$ satisfy:
\begin{subequations}
	\begin{align}
	\varepsilon(t) &= e^{(A-DC)(t-t_0)}\varepsilon(t_0) \nonumber\\
	& + \int_{\tau - t_0}^{t}e^{(A-DC)(t-\tau)}f_1n_1(\tau) \,\mathrm{d}t\\
	\varepsilon'(t) &= C\varepsilon(t)
	\end{align}
\end{subequations}

As can be observed from Section \ref{line_model}, when the error term $f_1$ is a constant vector and $n_1(t)$ is sinusoidal, the trajectory of the estimation error must be contained in the subspace spanned by $f_1$:
\begin{equation}
W = 
\begin{bmatrix}
f_1 & (A - DC)f_1 & \ldots & (A - DC)^{n-1}f_1
\end{bmatrix}
\end{equation}

The output error $\varepsilon'(t)$ is similarly constrained to the subspace spanned by $CW$.

As a result, it's desirable to design the feedback matrix $D$ such that for each fault event vector, the output error maintains a unique direction in the output space, meanwhile maintain the desired eigenvalues of $A-DC$. The detailed design in this paper relies on the concept of detection space, annihilating polynomial, cyclic invariant, detection generator, output separable, mutually detectable, excess subspace, output stationary and canonical form. The definitions can be referred in \cite{20}. In order to discuss the designing approach in \ref{filter}, some of the theorems from \cite{20} are briefly summarized below.

\newtheorem{theorem}{Theorem}
\begin{theorem} 
	\label{l1}
	Let
	\begin{equation}
	D_f = Af\left[ (Cf)^T(Cf) \right]^{-1}(Cf)^T
	\end{equation}
	and
	\begin{equation}
	C' = \left[ E_m - Cf\left[ (Cf)^T(Cf) \right]^{-1}(Cf)^T \right]C
	\end{equation}
	where $E_m$ is the identity matrix. Then the detection space is
	\begin{equation}
	\overline{R_f} = \eta(M_d')
	\end{equation}
	where 
	\begin{equation}
	M_d' = 
	\begin{bmatrix}
	C'\\
	C'(A - D_fC) \\
	\vdots \\
	C'(A-D_fC)^{n-1}
	\end{bmatrix}.
	\end{equation}
\end{theorem}

Theorem \ref{l1} provides a straightforward algorithm based on orthogonality for finding detection spaces for the event vector. It is possible that $\overline{R_f}$ is an invariant subspace with respect to $A-DC$. It is also possible to show that this is cyclic-invariant and there exists a unique vector $g$, defined as the \emph{detection generator}, in $\overline{R_f}$ such that the vectors $g, Ag, \ldots, A^{v-1}g$ are a basis for $\overline{R_f}$.

\begin{theorem}
	\label{l3}
	Let $g$ be the detection generator for $\overline{R_f}$. If $\psi_d(\cdot)$ is the $v_f$-order desired minimal polynomial for $\overline{R_f}$ with respect to $A-DC$, then:
	\begin{itemize}
		\item[i)] $A^kg \in R_f, \quad k < v_f - 1$.
		\item[ii)] $CA^{v_f-1}g = Cf$.
	\end{itemize}
	The statements i) and ii) imply that the vectors $g, Ag, \ldots, A^{v_f-1}g$ are a basis for $\overline{R_f}$.
	and $D$ must satisfy:
	\begin{equation}
	DCA^{v_f-1}g = \psi_d(A)g.
	\end{equation}
\end{theorem}
where $\psi_d(A)$ is the minimal annihilating polynomial of $g$ with respect to $A$.
Theorem \ref{l3} provides the numerical algorithm for computing the feedback matrix $D$ once all the detection spaces are defined.

\begin{theorem}
	\label{multiplefailuretheorem}
	The detection space $\overline{R_F}$ for multiple failure $F$ is the subspace defined similarly as the detection space for single failure subspace with
	\begin{subequations}
		\begin{align}
		\overline{R_F} &= \eta(M_D'), \\
		D_f &= AF\left[ (CF)^T(CF) \right]^{-1} (CF)^T, \\
		C_F' &= \left[ E_m - CF\left[ (CF)^T(CF) \right]^{-1}(CF)^T \right]C, \\
		M_D' &= 
		\begin{bmatrix}
		C_F' \\
		C_F'(A-D_FC) \\
		\vdots \\
		C_F'(A-D_FC)^{n-1}
		\end{bmatrix}.
		\end{align}
	\end{subequations}
\end{theorem}

\begin{theorem}
	\label{excessivesubspace}
	If $f_1, \ldots, f_r$ are output separable, then the relationship between individual detection spaces and the detection space $\overline{R_F}$ is:
	\begin{equation}
	\overline{R_1} \oplus \overline{R_2} \oplus \cdots \oplus \overline{R_r} \subset \overline{R_F}.
	\end{equation}
\end{theorem}

\begin{theorem}
	\label{mutualdetectable}
	The detection spaces $\overline{R_1}, \ldots, \overline{R_r}$ are mutually detectable if and only if
	\begin{equation}
	d(\overline{R_1}) + d(\overline{R_2}) + \cdots + d(\overline{R_r}) = d(\overline{R_F}).
	\end{equation}
\end{theorem}

\begin{theorem}
	\label{assignableeigenvalue}
	Let $\overline{R_0}$ be the excess subspace for $\overline{R_F}$ such that
	\begin{equation}
	\overline{R_F} = \overline{R_1} \oplus \cdots \oplus \overline{R_r} \oplus \overline{R_0}
	\end{equation}
	and
	\begin{equation}
	\overline{R_0} \subset \eta(C).
	\end{equation} 
	If $D$ is chosen such that $\overline{R_1}, \ldots, \overline{R_r}$ are invariant with respect to $A - DC$, then the $v_0 = d(\overline{R_0})$ eigenvalues of $A - DC$ associated with $\overline{R_0}$ are fixed and independent of the choice of $D$.
\end{theorem}

\begin{lemma}
	A \emph{canonical form} for the state space model can be achieved by coordinate transform from:
	\begin{equation}
	T^{-1} = 
	\begin{bmatrix}
	g_1 & \ldots & A^{v_1-1}g_1 & \ldots & g_r & \ldots & A^{v_r-1}g_r & T_0
	\end{bmatrix},
	\end{equation}
	where:
	\begin{equation}
	T_0 = 
	\begin{bmatrix}
	z_1 & \ldots & z_{v_0}
	\end{bmatrix},
	\end{equation}
	$z_1, \ldots, z_{v_0}$ is a basis for the excess subspace $R_0$.
	
	And the transformation for the output space is:
	\begin{equation}
	T_m^{-1} = 
	\begin{bmatrix}
	CA^{v_1-1}g_1 & \ldots & CA^{v_r-1}g_r
	\end{bmatrix}.
	\end{equation}
	
	The transformation to base normal form is:
	\begin{subequations}
		\begin{align}
		\hat{A} &= T^{-1}AT, \hat{B} = T^{-1}BT, \\
		\hat{C} &= T_m^{-1}CT,	\hat{D} = T^{-1}DT_m,
		\end{align}
	\end{subequations}
\end{lemma}

By the transformation, the selection of the feedback matrix $D$ does not have to concern the direction of the generated residual. However in this paper, the most significant advantage is that the transformation allows $\hat{A} - \hat{D}\hat{C}$ to be partially block diagonal. In this design approach since all detection spaces are non-intersecting and dimension 1, the generated residual for each fault event vector to would lie in independent dimensions as well.

\section{Designing Optimal Detection Filter}
\label{filter}

\subsection{bad data detection}
This design approach assigns all fault event vectors for direct detection of current measurement bad data:
\begin{equation}
\label{eq11}
F = 
\begin{bmatrix}
f_1 & f_2 & \ldots & f_8
\end{bmatrix}
=
\begin{bmatrix}
E_4 & 0 \\
0 & E_4 \\
0 & 0
\end{bmatrix}.
\end{equation}

And the canonical transformation allows each fault event vector $f_i$ only generates output residual in $e_i$, which is the $i_{th}$ basis. There are in total eight fault event vectors, the first four event vectors, $f_1$, $f_2$, $f_3$, and $f_4$ are assigned to detect the current measurement bad data through the left terminal, the second four event vectors, $f_5$, $f_6$, $f_7$, and $f_8$ are assigned to detect the current measurement bad data through the right terminal. 

According to Theorem \ref{l1}, with $C'$ and $D_f$ computed, the dimension of the observable subspaces $M_d'$ for all the eight event vectors $f_1, \ldots, f_8$ are found to be 11, i.e.,
\begin{equation}
d(M_d'(f_i)) = 11, \quad i = 1, 2, \ldots, 8.
\end{equation}

So that the dimensions of the null spaces of the observable subspaces are 1, i.e.,
\begin{equation}
v_i = d\left(\overline{R_i}\right) = d\left(\eta\left(M_d'\left(f_i\right)\right)\right), \quad i = 1, 2, \ldots, 8.
\end{equation}

And since $f_i \subset \bar{R_i}, \ i=1,2,\ldots,8$, it follows that without any further computation, the detection space for each fault vector is known to be spanned by the fault event vector itself. In matrix notation it means:
\begin{equation}
\bar{R_i} =
\begin{bmatrix}
f_i
\end{bmatrix}, \quad i = 1, 2, \ldots, 8.
\end{equation}

The computation of detection generators for $f_1, \ldots, f_8$ is also straightforward since from Theorem \ref{l3}
\begin{equation}
CA^{v_i - 1}g_i = Cf_i, \quad i = 1, 2, \ldots, 8.
\end{equation}

And since $v_i = 1$ for $i=1, 2, \ldots, 8$, it follows that the detection generator of each detection space is the fault event vector itself as well, i.e., $g_i=f_i,\ i=1,2,\ldots,8$.

According to Theorem \ref{multiplefailuretheorem}, the total detection space $\overline{R_F}$ for $F$ is the null space of $M_D'$, and since the dimension of the observable space:
\begin{equation}
d(M_D'(F)) = 0.
\end{equation}

where $M_D'$ is empty space, it implies that the detection space is full space:
\begin{equation}
\overline{R_F} = \delta^{12}.
\end{equation}

Since $d(F) = d(CF) = 8$, it follows that the fault event vectors are output separable. However from Theorem \ref{mutualdetectable}, since
\begin{equation}
\sum_{i=1}^8 d\left(\overline{R_i}\right) = 8 < d\left(\overline{R_F}\right) = 12.
\end{equation}
So the fault event vectors are not mutually detectable. Based on \ref{assignableeigenvalue} the excess subspace $R_0$ of dimension $v_0 = 4$ is needed such that:
\begin{equation}
\overline{R_F} = \overline{R_1} \oplus \cdots \oplus \overline{R_8} \oplus \overline{R_0}.
\end{equation}

And according to Theorem \ref{assignableeigenvalue}, four eigenvalues are independent of the choice of $D$, i.e., the four eigenvalues are fixed. Since the excess subspace belongs to the null space of $C$, i.e., $\overline{R_0} \subset \eta(C)$, and $d(\eta(C)) = 4$ as well as  $v_0 = 4$, the dimension of the  excess subspace is the same as the dimension of the null space of $C$, so that
\begin{equation}
\overline{R_0} = \eta(C).
\end{equation}

The general numerical computation of the excess subspace can be referred in \cite{20}. The procedure is to remove from $R_F$ the subspaces $R_1, \ldots, R_r$ in such a manner that only $R_0$ remains. 

Once the excess subspace and the unassignable eigenvalues of $A-DC$ have been determined, three options are available to the design:
\begin{itemize}
	\item[i)] Choose a $D$ for which each of $f_1, \ldots, f_8$ generate unidirectional output errors and accept the $v_0$ unassignable eigenvalues of $A-DC$.
	\item[ii)] Find a subset of $f_1, \ldots, f_8$ for which they are output separable and mutual detectable, and there are no unassignable eigenvalues.
	\item[iii)] Increase the dimension of the reference model such that all eigenvalues are assignable. 
\end{itemize}

In the simulation part of this paper, it is revealed that the unassignable eigenvalues are all negative. Thus option i) is acceptable for transmission line application.

Since only eight eigenvalues can be assigned arbitrarily, let the eigenvalues be $\lambda_i$ for detection space $\overline{R_i}, \ i=1,2,\ldots, 8$ respectively. The desired minimal annihilating polynomials are:
\begin{equation}
\psi_d(A)g_i = Ag-\lambda_ig_i, \quad i = 1,2, \ldots, 8.
\end{equation}

With the desired minimal annihilating polynomials for each detection space, the feedback matrix $D$ can be computed by Theorem \ref{l3}
\begin{equation}
D = 
\begin{bmatrix}
\psi_d(A)g_1 & \ldots & \psi_d(A)g_8
\end{bmatrix} \times
\begin{bmatrix}
Cg_1 & \ldots & Cg_8
\end{bmatrix}^{-1}.
\end{equation}

Finally the canonical form can be generated by the linear transformation using $T$ and $T_m$:
\begin{subequations}
	\begin{align}
	T^{-1} &= 
	\begin{bmatrix}
	g_1 & g_2 & \ldots & g_8 & z_1 & z_2 & z_3 & z_4
	\end{bmatrix}, \\
	T^{-1}_m &= 
	\begin{bmatrix}
	Cg_1 & Cg_2 & Cg_3 & Cg_4 & Cg_5 & Cg_6 & Cg_7 & Cg_8
	\end{bmatrix}.
	\end{align}
\end{subequations}
where the excess subspace is spanned by the linearly independent columns of $z_i$.
\begin{equation}
R_0 =
\begin{bmatrix}
z_1 & \ldots & z_{4}
\end{bmatrix}.
\end{equation}

The canonical form allows the fault event vector $f_i$ to generate the output residual only in $e_i, \ i=1,2,\ldots,8$, which is the $i_{th}$ basis vector.

\subsection{fault detection}
From the design of the fault event vectors, it can be seen that single phase to ground faults can be represented as linear combinations of the fault event vectors. 

Take the single phase to ground fault for example, since the fault residual is:
\begin{equation}
f_{\textnormal{A-G}}(t) = \left(\left(1-\alpha\right)f_1 + \alpha f_5\right)\frac{v_{af}(t)}{R_f}.
\end{equation}

The actual estimation error in canonical form is computed as:
\begin{equation}
\hat{\varepsilon}_{\textnormal{A-G}}(t) = \int_{\tau = t_0}^{t} e^{\left(\hat{A}-\hat{D}\hat{C}\right)(t-\tau)}T^{-1}f_{\textnormal{A-G}}(\tau)\,\mathrm{d}\tau
\end{equation}
by assuming $\varepsilon(t_0) = 0$ as all residuals grow exponentially to 0 by the negative value of assigned eigenvalues. $\hat{\varepsilon}$ is the estimation error after canonical transformation $T$.

And the output residual in canonical form is:
\begin{subequations}
	\begin{align}
	\hat{\varepsilon}_{\textnormal{A-G}}'(t) &= \hat{C}\hat{\varepsilon}_{\textnormal{A-G}}(t) \\
	&= (1-\alpha)\int_{\tau = t_0}^{t} \hat{C}e^{\left(\hat{A}-\hat{D}\hat{C}\right)(t-\tau)}T^{-1}f_1(\tau)\frac{v_{af}(\tau)}{R_f}\,\mathrm{d}\tau \nonumber\\
	&+ \alpha\int_{\tau = t_0}^{t}\hat{C}e^{\left(\hat{A}-\hat{D}\hat{C}\right)(t-\tau)}T^{-1}f_5(\tau)\frac{v_{af}(\tau)}{R_f}\,\mathrm{d}\tau \\
	&= (1-\alpha)\int_{\tau = t_0}^{t} \hat{C}e^{\left(\hat{A}-\hat{D}\hat{C}\right)(t-\tau)}T^{-1}\frac{v_{af}(\tau)}{R_f}\,\mathrm{d}\tau f_1 \nonumber\\
	&+ \alpha\int_{\tau = t_0}^{t}\hat{C}e^{\left(\hat{A}-\hat{D}\hat{C}\right)(t-\tau)}T^{-1}\frac{v_{af}(\tau)}{R_f}\,\mathrm{d}\tau f_5 \\
	&= (1-\alpha)S_{\textnormal{A-G}}(t)f_1 + \alpha S_{\textnormal{A-G}}(t)f_5.
	\end{align}
\end{subequations}

The derivation comes from the fact that $f_1$ and $f_2$ are only constant vectors and they can be pulled outside the integral. Before this operation the integral is performed on a vector, and after this operation the integral is performed on a matrix. Specifically, the matrix is: 
\begin{equation}
S_{\textnormal{A-G}}(t) = \int_{\tau = t_0}^{t} \hat{C}e^{\left(\hat{A}-\hat{D}\hat{C}\right)(t-\tau)}T^{-1}\frac{v_{af}(\tau)}{R_f}\,\mathrm{d}\tau.
\end{equation}

Since by the choice of the feedback matrix $D$ and the coordinate transformation, it is guaranteed that the integral upon $f_1(t)$ lies only in the direction of $e_1$ and the integral upon $f_2(t)$ lies only in the direction of $e_2$, the product of the time varying matrix with the fault event vector can be represented by the product of a scalar with the fault event vector:
\begin{subequations}
	\begin{align}
	S_{\textnormal{A-G}}(t)f_1 &= s_{\textnormal{A-G}}(t)e_1, \\
	S_{\textnormal{A-G}}(t)f_5 &= s_{\textnormal{A-G}}(t)e_5.
	\end{align}
\end{subequations}

So that the combined output residual is:
\begin{equation}
\hat{\varepsilon}_{\textnormal{A-G}}'(t) = 
\begin{bmatrix}
(1-\alpha)s_{\textnormal{A-G}}(t) & 0_{1\times 3} & \alpha s_{\textnormal{A-G}}(t) & 0_{1\times 7}
\end{bmatrix}^T.
\end{equation}

For the same reason, a single phase B to ground fault would generate a similar output residual as:
\begin{equation}
\hat{\varepsilon}_{\textnormal{B-G}}'(t) = 
\begin{bmatrix}
0 & (1-\alpha)s_{\textnormal{B-G}}(t) & 0_{1\times 3} & \alpha s_{\textnormal{B-G}}(t) & 0_{1\times 6}
\end{bmatrix}^T.
\end{equation}

A single phase C to ground fault would generate a similar output residual as:
\begin{equation}
\hat{\varepsilon}_{\textnormal{C-G}}'(t) = 
\begin{bmatrix}
0_{1\times 2} & (1-\alpha)s_{\textnormal{C-G}}(t) & 0_{1\times 3} & \alpha s_{\textnormal{C-G}}(t) & 0_{1\times 5}
\end{bmatrix}^T.
\end{equation}

To conclude, for any single phase to ground fault, the output residuals are only non-zero in two directions: phase A-ground, B-ground and C-ground faults generate output residual in $e_1$/$e_5$, $e_2$/$e_6$, and $e_3$/$e_7$. This also proves that the single phase to ground faults are output stationary to $f_1, \ldots, f_8$. Also, the fault location as represented in percentage can be computed as ratio between the magnitudes of the two non-zero output residuals. For example, in single phase A to ground fault, the fault location can be computed by:
\begin{equation}
\frac{1-\alpha}{\alpha} = \frac{\hat{\varepsilon}_{\textnormal{A-G}}'(t,1)}{\hat{\varepsilon}_{\textnormal{A-G}}'(t,5)}.
\end{equation}
where $\hat{\varepsilon}_{\textnormal{A-G}}'(t,i)$ is the $i$th row of transformed output residual at time $t$. The fault location is with respect to the left terminal of the transmission line.

On the other hand, the phase to phase fault behaves similarly. Take an example for the phase A to phase B fault, the fault residual is:
\begin{equation}
f_{\textnormal{A-B}}(t) = \left( (1-\alpha)(f_1 - f_2) + \alpha(f_5 - f_6)\right) \frac{v_{af}(t) - v_{bf}(t)}{R_f}.
\end{equation}

The output residual in canonical form is:
\begin{multline}
	\hat{\varepsilon}_{\textnormal{A-B}}'(t) = \hat{C}\hat{\varepsilon}_{\textnormal{A-B}}(t) \\
	= (1-\alpha)\int_{\tau = t_0}^{t} \hat{C}e^{\left(\hat{A}-\hat{D}\hat{C}\right)(t-\tau)}T^{-1}  \frac{v_{af}(\tau)-v_{bf}(\tau)}{R_f}\,\mathrm{d}\tau f_1 \\
	-(1-\alpha)\int_{\tau = t_0}^{t} \hat{C}e^{\left(\hat{A}-\hat{D}\hat{C}\right)(t-\tau)}T^{-1}  \frac{v_{af}(\tau)-v_{bf}(\tau)}{R_f}\,\mathrm{d}\tau f_2 \\
	+ \alpha\int_{\tau = t_0}^{t} \hat{C}e^{\left(\hat{A}-\hat{D}\hat{C}\right)(t-\tau)}T^{-1}  \frac{v_{af}(\tau)-v_{bf}(\tau)}{R_f}\,\mathrm{d}\tau f_5 \\
	-\alpha\int_{\tau = t_0}^{t} \hat{C}e^{\left(\hat{A}-\hat{D}\hat{C}\right)(t-\tau)}T^{-1} \frac{v_{af}(\tau)-v_{bf}(\tau)}{R_f}\,\mathrm{d}\tau f_6 \\
	= (1-\alpha)S_{\textnormal{A-B}}(t)f_1 - (1-\alpha) S_{\textnormal{A-B}}(t)f_2 \\
	+ \alpha S_{\textnormal{A-B}}(t)f_5 - \alpha S_{\textnormal{A-B}}(t)f_6.
\end{multline}

Or,
\begin{multline}
\hat{\varepsilon}_{\textnormal{A-B}}'(t) =
\big[ (1-\alpha)s_{\textnormal{A-B}}(t),  -(1-\alpha)s_{\textnormal{A-B}}(t), 0, 0, \\
\alpha s_{\textnormal{A-B}}(t), -\alpha s_{\textnormal{A-B}}(t), 0_{1\times 6} \big]^T.
\end{multline}

Similarly,
\begin{multline}
\hat{\varepsilon}_{\textnormal{B-C}}'(t) =
\big[0, (1-\alpha)s_{\textnormal{B-C}}(t),  -(1-\alpha)s_{\textnormal{B-C}}(t), 0, 0, \\
\alpha s_{\textnormal{B-C}}(t), -\alpha s_{\textnormal{B-C}}(t), 0_{1\times 5} \big]^T.
\end{multline}
\begin{multline}
\hat{\varepsilon}_{\textnormal{C-A}}'(t) =
\big[-(1-\alpha)s_{\textnormal{C-A}}(t), 0, (1-\alpha)s_{\textnormal{C-A}}(t), 0, \\
-\alpha s_{\textnormal{C-A}}(t), 0, \alpha s_{\textnormal{C-A}}(t), 0_{1\times 5} \big]^T.
\end{multline}

The fault location in percentage for phase A-B fault can either be computed by ratio between the output residual in $f_1$ and the output residual in $f_5$, or between the output residual in $f_2$ and the output residual in $f_6$.

Finally the three phase fault residual is:
\begin{multline}
	f_{\textnormal{A-B-C}}(t) = \left((1-\alpha)f_1 + \alpha f_5\right) \frac{2v_{af}(t) - v_{bf}(t) - v_{cf}(t)}{R_f} \\
	= \left((1-\alpha)f_2 + \alpha f_6\right) \frac{2v_{bf}(t) - v_{cf}(t) - v_{af}(t)}{R_f} \\
	= \left((1-\alpha)f_3 + \alpha f_7\right) \frac{2v_{cf}(t) - v_{af}(t) - v_{bf}(t)}{R_f}
\end{multline}

The output residual is:
\begin{multline}
\hat{\varepsilon}_{\textnormal{A-B-C}}'(t) = \big[ (1-\alpha)s_{\textnormal{A-B-C-1}}(t), (1-\alpha)s_{\textnormal{A-B-C-2}}(t), \\ 
(1-\alpha)s_{\textnormal{A-B-C-3}}(t), 0, \alpha s_{\textnormal{A-B-C-1}}(t), \\ \alpha s_{\textnormal{A-B-C-2}}(t), \alpha s_{\textnormal{A-B-C-3}}(t), 0_{1\times 5} \big]^T
\end{multline}
where
\begin{subequations}
	\begin{align}
	S_{\textnormal{A-B-C-1}}(t) &= \int_{\tau = t_0}^{t} \hat{C}e^{\left(\hat{A}-\hat{D}\hat{C}\right)(t-\tau)}T^{-1} \nonumber\\
	& \frac{2v_{af}(\tau) - v_{bf}(\tau) - v_{cf}(\tau)}{R_f}\,\mathrm{d}\tau, \\
	S_{\textnormal{A-B-C-1}}(t)f_1 &= s_{\textnormal{A-B-C-1}}(t)f_1, \\
	S_{\textnormal{A-B-C-1}}(t)f_5 &= s_{\textnormal{A-B-C-1}}(t)f_5, \\
	S_{\textnormal{A-B-C-2}}(t) &= \int_{\tau = t_0}^{t} \hat{C}e^{\left(\hat{A}-\hat{D}\hat{C}\right)(t-\tau)}T^{-1} \nonumber\\
	& \frac{2v_{bf}(\tau) - v_{cf}(\tau) - v_{af}(\tau)}{R_f}\,\mathrm{d}\tau, \\
	S_{\textnormal{A-B-C-2}}(t)f_2 &= s_{\textnormal{A-B-C-2}}(t)f_2, \\
	S_{\textnormal{A-B-C-2}}(t)f_6 &= s_{\textnormal{A-B-C-2}}(t)f_6, \\
	S_{\textnormal{A-B-C-3}}(t) &= \int_{\tau = t_0}^{t} \hat{C}e^{\left(\hat{A}-\hat{D}\hat{C}\right)(t-\tau)}T^{-1} \nonumber\\
	& \frac{2v_{cf}(\tau) - v_{af}(\tau) - v_{bf}(\tau)}{R_f}\,\mathrm{d}\tau, \\
	S_{\textnormal{A-B-C-3}}(t)f_3 &= s_{\textnormal{A-B-C-3}}(t)f_3, \\
	S_{\textnormal{A-B-C-3}}(t)f_7 &= s_{\textnormal{A-B-C-3}}(t)f_7.
	\end{align}
\end{subequations}

The conclusion from the discussion gives the following theorem:
\begin{theorem}
	With the fault event vector designed as in (\ref{eq11}) targeting at bad current measurement, the internal faults can be detected by non-zero output residuals in corresponding phase, and the fault location $\alpha (0<\alpha<1)$ can be computed as the ratio of the non-zero output residuals from the corresponding faulted phase.
\end{theorem}

\section{Simulation Result}
\label{simulation}
The test case is a two bus system with the detection filter monitoring a 128 km long line as illustrated in Fig. 1. The parameters of the transmission line are summarized in Table \ref{table1}. The rated voltage is 115 kV and the rated current is 1300 A.

\begin{table}[!t]
	\renewcommand{\arraystretch}{1.3}
	\caption{Transmission Line Parameters}
	\label{table1}
	\centering
	\begin{tabular}{|c|c|c|c|c|}
		\hline
		$R$ ($\mathrm{\Omega}$) & A & B & C & N\\
		\hline
		A & 12.270 & 7.180 & 7.197 & 6.748 \\
		\hline
		B & 7.180 & 12.310 & 7.216 & 6.750 \\
		\hline
		C & 7.197 & 7.216 & 12.350 & 6.757 \\
		\hline
		N & 6.748 & 6.750 & 6.757 & 147.3 \\
		\hline
		$L$ ($\mathrm{H}$) & A & B & C & N\\
		\hline
		A & 0.2881 & 0.1521 & 0.1342 & 0.1472 \\
		\hline
		B & 0.1521 & 0.2878 & 0.1519 & 0.1331 \\
		\hline
		C & 0.1342 & 0.1519 & 0.2878 & 0.1234 \\
		\hline
		N & 0.1472 & 0.1331 & 0.1234 & 0.4356 \\
		\hline
		$\mathrm{Cap}$ ($\mu$F) & A & B & C & N\\
		\hline
		A & 0.5624 & -0.1447 & -0.0728 & -0.1086 \\
		\hline
		B & -0.1447 & 0.5807 & -0.1479 & -0.0565 \\
		\hline
		C & -0.0728 & -0.1479 & 0.5525 & -0.0387 \\
		\hline
		N & -0.1086 & 0.1331 & -0.0387 & 0.4104 \\
		\hline
	\end{tabular}
\end{table}

\begin{figure}[!t]
	\centering
	\includegraphics[width=3.5in]{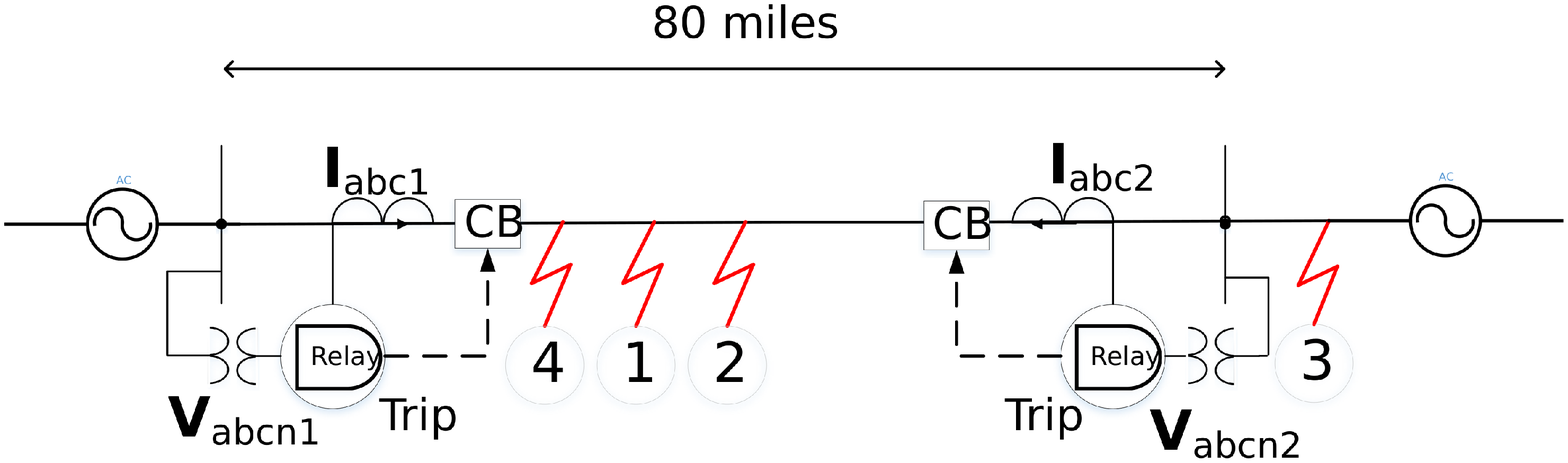}
	\caption{XXX Need to change this to WINIGS model.}
	\label{fig1}
\end{figure}

The simulated events are summarized in Table \ref{table2}. The monitored transmission line is operating normally from time $t=0.0 \mathrm{s}$ to $0.5\mathrm{s}$. Subsequently different single phase to ground faults, phase to phase faults and three phase faults with various fault resistance and fault location are simulated. The column of fault location indicates the distance of the fault to the left terminal of the line. The available measurements are all phase current measurements on both sides of the terminals, and three phase to neutral voltage measurements on both sides of the terminals. The neutral to ground voltage measurement is assumed as pseudo measurement with a constant value of $0$. All current measurements are treated in the state space model as system control inputs, while all voltage measurements are treated in the state space as system output.

\begin{table}[!t]
	\renewcommand{\arraystretch}{1.3}
	\caption{Simulated Events for Test System}
	\label{table2}
	\centering
	\begin{tabular}{|c|c|c|c|c|c|}
		\hline
		Time & Event & Event  & Fault & Fault & Internal \\
		 ($\mathrm{s}$) & \# & Type & Resistance ($\mathrm{\Omega}$) & Location ($\mathrm{km}$) & or External \\
		\hline
		0.1--0.5 & 0 & N/A & N/A & N/A & N/A \\
		\hline
		0.6--0.8 & 1 & A-G & 1000 & 48 & Internal \\
		\hline
		1.0--1.2 & 2 & B-G & 500 & 48 & Internal \\
		\hline
		1.4--1.6 & 3 & B-C & 0.5 & 48 & Internal \\
		\hline
		1.8--2.0 & 4 & C-G & 500 & 64 & Internal \\
		\hline
		2.2--2.4 & 5 & A-C & 10 & 64 & Internal \\
		\hline
		2.6--2.8 & 6 & A-B & 20 & 64 & Internal \\
		\hline
		3.0--3.2 & 7 & A-B-C & 1 & 128.032 & External \\
		\hline
		3.4--3.6 & 8 & A-B-C & 2 & 16 & Internal \\
		\hline
		3.8--4.0 & 9 & A-G & 1 & 16 & Internal \\
		\hhline{|=|=|=|=|=|=|}
		Time & Event & Event & \multirow{2}{*}{Bad Data} & Bad Data &  \\
		 ($\mathrm{s}$) & \# & Type & & Type & \\
		\hline
		4.2--4.4 & 10 & A & Left terminal & Loss of current & \\
		\hline
		4.6--4.8 & 11 & B & Left terminal & Loss of current & \\
		\hline
		5.0--5.2 & 12 & C & Left terminal & Loss of current & \\
		\hline
		5.4--5.6 & 13 & A & Right terminal & Loss of current & \\
		\hline
		5.8--6.0 & 14 & B & Right terminal & Loss of current & \\
		\hline
		6.2--6.4 & 15 & C & Right terminal & Loss of current & \\
		\hline
	\end{tabular}
\end{table}

The unassignable eigenvalues computed numerically are 0.8118, 0.9940, 0.9957, 0.9949 respectively. The assignable eigenvalues are set at 0.1. The measurement noise for each channel is within 0.02 p.u. Such level is used as threshold to determine if non-zero residual is caused by noise or by fault.

\begin{figure}[!t]
	\centering
	\includegraphics[width=3.5in]{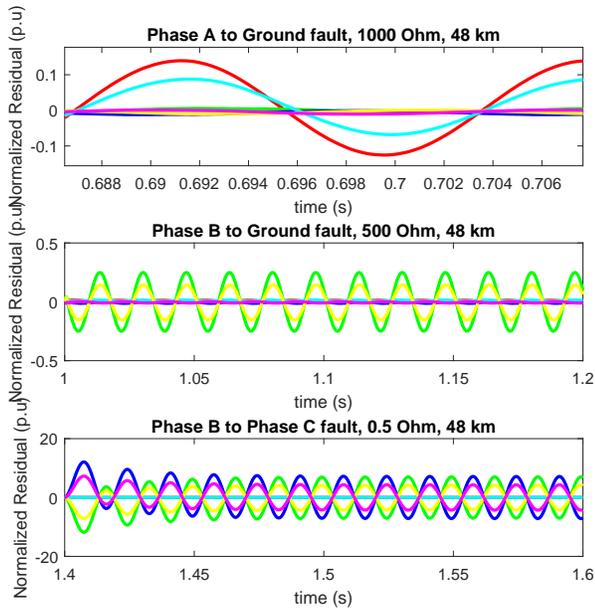}
	\caption{Output Residuals for Faults at 48km, Corresponding to Event 1, 2 and 3}
	\label{fig4Residual48Km}
\end{figure}

\begin{figure}[!t]
	\centering
	\includegraphics[width=3.5in]{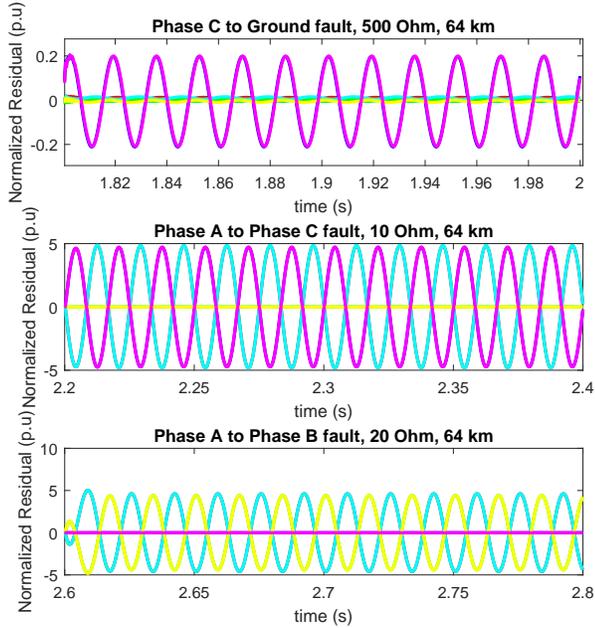}
	\caption{Output Residual for Faults at 64km.}
	\label{fig5Residual64Km}
\end{figure}

\begin{figure}[!t]
	\centering
	\includegraphics[width=3.5in]{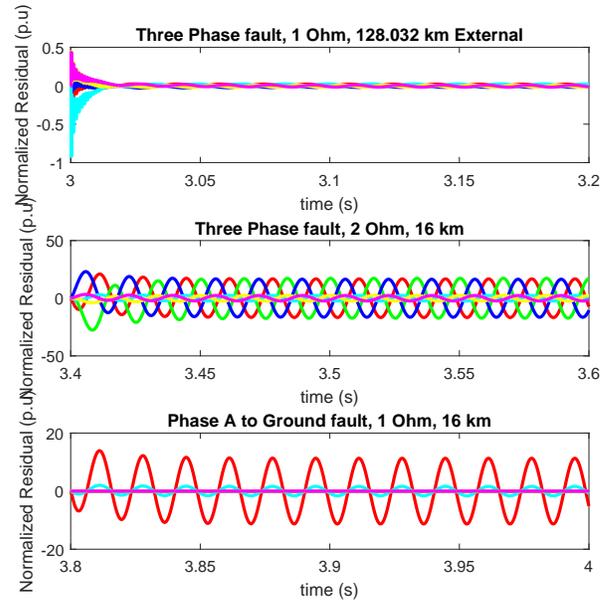}
	\caption{Output Residual for External and Faults at 16km.}
	\label{fig6Residual16Km}
\end{figure}

\begin{figure}[!t]
	\centering
	\includegraphics[width=3.5in]{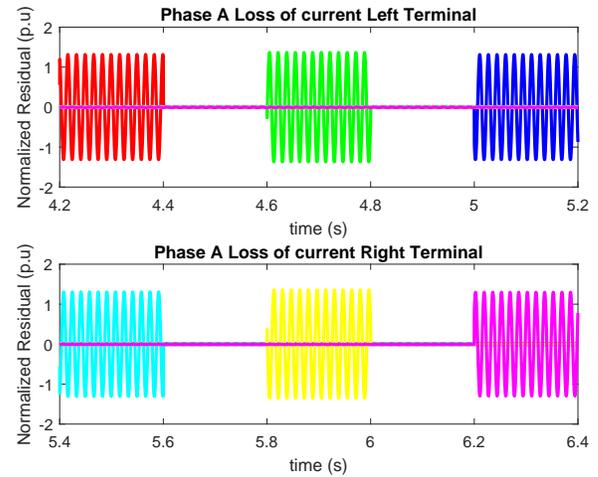}
	\caption{Output Residual for Loss of Current Measurement.}
	\label{fig7LOC}
\end{figure}

The filtering results are presented from Figs \ref{fig4Residual48Km} to \ref{fig7LOC}. For each graph there are six traces. The red/green/blue traces are the residuals generated for phase A/B/C on the left terminal of the line, the cyan/yellow/magenta traces are the residuals generated for phase A/B/C on the right terminal of the line.

The residuals generated for the internal faults at $48 \,\mathrm{km}$ from left terminal of the line are illustrated in Fig. \ref{fig4Residual48Km}. For the phase A to ground high impedance fault, only two residuals have magnitude larger than $0.02 \,\mathrm{pu}$---the residual for phase A left terminal is a sinusoidal wave with maximum magnitude of $0.1393 \,\mathrm{pu}$, while the residual for phase A right terminal has maximum magnitude of $0.0877 \,\mathrm{pu}$. The computed fault location is $46.63 \,\mathrm{km}$. The 1.07\% computation error is from the modeling error. The fault location error would decrease when a more severe fault generates larger residual. 

For the phase B to ground high impedance fault, the residual for phase B left terminal has maximum magnitude of $0.2618 \,\mathrm{pu}$, the residual for phase B right terminal has maximum magnitude of $0.152 \,\mathrm{pu}$. The computed fault location is $47.018 \,\mathrm{km}$. The 0.7\% computation error comes from the modeling error. 

When the fault is more severe the modeling error would contribute less to the fault location inaccuracy, this is illustrated for the phase B to C fault. Only four residuals have maximum magnitude larger than $0.02 \,\mathrm{pu}$---the residual for phase B left terminal has maximum magnitude of $7.322 \,\mathrm{pu}$, while the residual for phase B right terminal has maximum magnitude of $4.422 \,\mathrm{pu}$ (and residuals for phase C). The computed fault location is 48.15 km. This accounts for only 0.1\% error.

The residuals generated for the internal faults at $64 \,\mathrm{km}$ (50\% of the line) are illustrated in Fig. \ref{fig5Residual64Km}. Similarly for single phase to ground fault, two residuals have magnitude exceeding the threshold, and for phase to phase faults, four residuals have magnitude exceeding the threshold. For the single phase C to ground fault, the phase A to C fault, and the phase A to B fault, the computed fault locations for all three cases are 0.13\% error. 

The residuals generated for three phase external faults, three phase internal faults and single phase A to ground faults at $16 \,\mathrm{km}$ from left terminal are illustrated in Fig. \ref{fig6Residual16Km}. For three phase internal faults, phase A/B/C has residual of magnitude 16.77 p.u (left) and 2.421 p.u (right), which accounts for 16.15 km or 0.12\% error. 

On the other hand, when there is current measurement bad data, only one residual corresponding to the bad data channel would have magnitude larger than the threshold, this is illustrated in Fig \ref{fig7LOC}.

\section{Conclusion}
This paper addresses the general fault diagnosis problem of power transmission network. The contribution of this paper includes 1) it proves that the internal faults of transmission line and bad data are equivalent and can be written in the same control failure form, and it proposes to use the linear observer theory to map each fault or bad data to a certain residual channel, 2) it provides one design approach for transmission line monitoring that allows detection of all types of internal fault, bad data detection for current measurements, and fault location. 

Further work remains for this research, typically signature CT saturation with bolted fault.


%

%
%
%
%
%

\ifCLASSOPTIONcaptionsoff
  \newpage
\fi

\begin{IEEEbiography}	[{\includegraphics[width=1in,height=1.25in,clip,keepaspectratio]{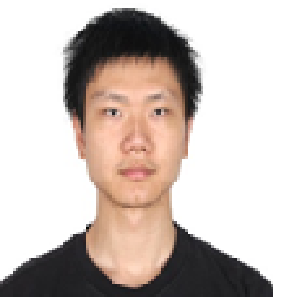}}]{Zhenyu Tan} (S'12) received B.E. degree in Electrical Engineering from Tsinghua University in Beijing, China in 2011. He received Ph.D degree Power System Protection area, in the Power System Control and Automation Laboratory from Georgia Institute of Technology, Atlanta, GA, USA.  
\end{IEEEbiography}
\begin{IEEEbiography}
	[{\includegraphics[width=1in,height=1.25in,clip,keepaspectratio]{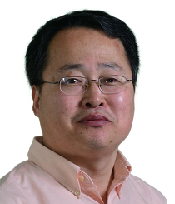}}]{Hongbo Sun} (SM'00) received his Ph.D. from  in Electrical Engineering from Chongqing University, China in 1991. He is currently a Senior Principal Research Scientist at Mitsubishi Electric Research Laboratories(MERL), Cambridge, Massachusetts. Prior to joining MERL, Hongbo was a principal application engineer at Oracle, and a technical arthitect at SPL WorldGroup. He is a senior member of IEEE. His research interest include power system modeling and analysis, power system operation and control, and smart grid applications.
\end{IEEEbiography}

\end{document}